\documentclass[12pt]{article}

\newcommand{\beq}{\begin{equation}}
\newcommand{\eeq}{\end{equation}}
\newcommand{\beqa}{\begin{eqnarray}}
\newcommand{\eeqa}{\end{eqnarray}}
\newcommand{\beqar}{\begin{eqnarray*}}
\newcommand{\eeqar}{\end{eqnarray*}}

\newcommand\bu{\bar{u}}

\newcommand{\dga}{\dot{\gamma}}
\newcommand{\dde}{\dot{\delta}}
\newcommand{\al}{\alpha}
\newcommand{\be}{\beta}

\def\spa          {\ \ \ }
\def\non          {\nonumber}
\def\ha           {\mbox{$\frac{1}{2}$}}

\def\spa          {\ \ \ }
\def\mand         {\spa\mbox{and}\spa}

\def\Tr           {\mbox{\rm Tr}\,}

\def\cd           {{\cdot}}
\def\ran          {\rangle}
\def\lan          {\langle}
\def\fsC    {C\!\!\!\!/\,}
\def\fsH    {H\!\!\!\!/\,}
\newcommand{\del}{\delta}

\newcommand{\eps}{\epsilon}
\newcommand{\ga}{\gamma}
\newcommand{\Ga}{\Gamma}

\newcommand{\inn}{\!\cdot\!}

\newcommand{\lam}{\lambda}

\newcommand{\z}{\zeta}

\newcommand{\labell}[1]{\label{#1}} 
\newcommand{\reef}[1]{(\ref{#1})}

\def\sst#1{{\scriptscriptstyle #1}}
\def\0{{\sst{(0)}}}
\def\1{{\sst{(1)}}}
\def\2{{\sst{(2)}}}
\def\3{{\sst{(3)}}}
\def\4{{\sst{(4)}}}
\def\5{{\sst{(5)}}}
\def\6{{\sst{(6)}}}
\def\7{{\sst{(7)}}}
\def\8{{\sst{(8)}}}

\begin{document}
\baselineskip 18pt%
\begin{titlepage}
\vspace*{1mm}%
 
 TUW-16-22

\vspace*{8mm}
\vspace*{8mm}%

\center{ {\bf \Large  On WZ and RR Couplings of BPS Branes \\
and their all order $\alpha'$ Corrections in  IIB, IIA
   }} 

\begin{center}
{Ehsan Hatefi   }\footnote{
ehsan.hatefi@tuwien.ac.at, ehsan.hatefi@cern.ch, e.hatefi@qmul.ac.uk, ehsanhatefi@gmail.com }

\vspace*{0.6cm}{ Institute for Theoretical Physics, TU Wien
\\
Wiedner Hauptstrasse 8-10/136, A-1040 Vienna, Austria}

\end{center}
\begin{center}{\bf Abstract}\end{center}
\begin{quote}

We compute all three and four point couplings of BPS $D_{p}$-branes for all different nonzero $p$-values on the entire world volume and transverse directions. We start finding out all four point function supersymmetric Wess-Zumino (WZ) actions of one closed string Ramond-Ramond (RR) field  with two fermions, either with the same (IIB) or different chirality (IIA) as well as their all order $\alpha'$ corrections. The closed form of S-matrices of two closed string RR in both IIB, IIA, including their all order $\alpha'$ corrections have also been addressed. Our results confirm  that, not only the structures of $\alpha'$ corrections but also their coefficients of IIB are quite different from their IIA ones.

The S-matrix of an RR and two gauge (scalar) fields and their all order corrections in  antisymmetric picture of RR  have been carried out as well. Various remarks on the restricted Bianchi identities as well as all order $\alpha'$ corrections to all different supersymmetric WZ couplings in both type IIA and IIB superstring theory are also released. Lastly, different singularity structures as well as all order contact terms for all non-vanishing traces in type II have  also been constructed.

 \end{quote}
\end{titlepage}

 \section{Introduction}
Dp-branes have been taken to be the known sources for Ramond-Ramond (RR) closed string
for all kinds of BPS branes  \cite{Polchinski:1995mt,Witten:1995im}. RR couplings have been playing the major roles in many areas , for instance the phenomenon dissolving  brane within branes \cite{Douglas:1995bn} , K-theory (through D-brane approach)
\cite{Minasian:1997mm,Witten:1998cd}, Myers effect \cite{Myers:1999ps} and some of their $\alpha'$ corrections \cite{Howe:2006rv,Hatefi:2012zh} are revealed. To describe the dynamics of D-branes, their effective actions must be established and almost all relevant literatures have been pointed out in the  introduction of \cite{Hatefi:2016wof}.  By dealing with Conformal Field Theory (CFT) and evaluating scattering amplitudes, we hope to enhance our knowledge of knowing new Effective Actions, more crucially, given S-Matrix formalism, among the other applications to this formalism,  new approaches to Effective Field Theory (EFT) will be discovered. In this note, we just highlight some of the applications that have been worked out such as  $N^3$ phenomenon for$M5$ branes, dS solutions and the entropy growth \cite{Hatefi:2012bp}. RR Couplings with non-BPS branes have also been figured out  by \cite{Kennedy:1999nn,Michel:2014lva} and the three ways of deriving EFT couplings are clarified in detail \cite{Hatefi:2012wj}. As last remark, we emphasize that by just going through S-matrix calculations, not only one is able to  construct new string couplings  but also exactly gain the coefficients of all the  higher derivative corrections to all orders in $\alpha'$.
One may find out some partial results of BPS string amplitude computations in \cite{Fotopoulos:2001pt}.
\vskip.2in

Here is the outline of the paper. We first  explore the supersymmetric Wess-Zumino (WZ)couplings of an RR and two fermion fields with either the same or different chirality of both IIB and IIA and then start building their all order $\alpha'$ corrections. Our computations clarify that
not only the structures of $\alpha'$ corrections but also their coefficients of IIB are quite different from their IIA ones. We then deal with all symmetric and asymmetric amplitudes of three and four point functions of an RR and two gauge (scalar) fields and reconstruct their all order corrections as well. 

\vskip.1in

 In \cite{Sen:2015hia} the role of picture-changing operators for perturbative string computations has been discussed, more importantly, in section three the whole setup was made. It was also argued that to calculate scattering amplitudes at each order  not only one must take into account all pieces together with local descriptions but also one needs to consider the vertical integration method which actually avoids all spurious singularities and a clear example was given in section 3.2 as well. Potentially its analysis  has something to do with disc string computations, however, in \cite{Sen:2015hia} it is not discussed how to 
find out RR bulk momenta.  Note that the correlation function  $<\partial^i X(x_{1}) e^{ip.x(z)}  >$ (between scalar field vertex operator in zero picture and exponential part of RR vertex operator) has non-zero contribution to our S-matrices and therefore there will be non-zero terms in S-matrices such as $p.\xi_1$ and  $p.\xi_2$ terms. These terms are related to  RR bulk momenta as they clearly carry momentum of RR in transverse directions. Indeed,  unlike \cite{Sen:2015hia}, in this paper we clearly  keep track of all S-matrix elements, including the terms that carry RR's momenta in the bulk, for instance $p.\xi_1$ and  $p.\xi_2$ terms. 

\vskip.1in

To make sense of all supersymmetric WZ actions  in antisymmetric picture of RR, we make various remarks on the restricted Bianchi identities. Ultimately all contact terms for various field content and their all order $\alpha'$ corrections for different WZ couplings in both type IIA and IIB and for all non-vanishing traces will be constructed out.

\section{The $\bar\Psi-\Psi-C$ of type IIB}

In this section we  would like to directly apply all the  CFT techniques \cite{Friedan:1985ge} to  actually derive entirely all supersymmetric WZ actions including their all order $\alpha'$ corrections. All  four point functions of a closed string RR and two fermion vertex operators with either the same or different chirality in ten dimensional space time of type IIB (IIA) superstring theory are going to be explored.  Hence, this four point function in IIB of BPS $D_{p}$-branes is given by the following  correlation function 
\beqa
{\cal A}^{\bar\Psi\Psi RR} & \sim & \int dx_1dx_2d^2z
 \lan
V_{\bar\Psi}^{(-1/2)}(x_1)V_{\Psi}^{(-1/2)}(x_2)
V_{RR}^{(-1)}(z,\bar{z})\ran,\labell{cor1}\eeqa
where  all the vertex operators are given by 
\beqa
V_{\bar\Psi}^{(-1/2)}(x)&=&\bar u^{\gamma}e^{-\phi(x)/2}S_{\gamma}(x)\,e^{ \alpha'iq.X(x)} \nonumber\\
V_{\Psi}^{(-1/2)}(x)&=&u^{\delta}e^{-\phi(x)/2}S_{\delta}(x)e^{ \alpha'  iq.X(x)}\nonumber\\
V_{C}^{(-\frac{1}{2},-\frac{1}{2})}(z,\bar{z})&=&(P_{-}\fsH_{(n)}M_p)^{\alpha\beta}e^{-\phi(z)/2}
S_{\al}(z)e^{i\frac{\alpha'}{2}p\cd X(z)}e^{-\phi(\bar{z})/2} S_{\be}(\bar{z})
e^{i\frac{\alpha'}{2}p\cd D \cd X(\bar{z})},\nonumber\eeqa
\vskip .2in
Our notation is such that $\mu,\nu=0,1,..,9$, world volume directions are shown by $a,b,c=0,1,...,p$ and transverse indices are give by $i,j=p+1,...,9$.
 All the objects are massless, where the  projector, RR's field strength  and spinor are shown by
\begin{displaymath}
P_{-} =\ha (1-\ga^{11}), \quad
\fsH_{(n)} = \frac{a
_n}{n!}H_{\mu_{1}\ldots\mu_{n}}\ga^{\mu_{1}}\ldots
\ga^{\mu_{n}}, (P_{-}\fsH_{(n)})^{\al\be} =
C^{\al\del}(P_{-}\fsH_{(n)})_{\del}{}^{\be}.
\non\end{displaymath}

For type IIA  (type IIB) $n=2,4$,$a_n=i$  ($n=1,3,5$,$a_n=1$). We would like to deal with just the holomorphic components of the world-sheet fields, therefore,  we are going to employ the so called doubling trick, that is, the following change of variables have to be taken into consideration
\begin{displaymath}
\tilde{X}^{\mu}(\bar{z}) \rightarrow D^{\mu}_{\nu}X^{\nu}(\bar{z}) \ ,
\spa
\tilde{\psi}^{\mu}(\bar{z}) \rightarrow
D^{\mu}_{\nu}\psi^{\nu}(\bar{z}) \ ,
\spa
\tilde{\phi}(\bar{z}) \rightarrow \phi(\bar{z})\,, \mand
\tilde{S}_{\al}(\bar{z}) \rightarrow M_{\al}{}^{\be}{S}_{\be}(\bar{z})
 \ ,
\non\end{displaymath}

with the following matrices
\begin{displaymath}
D = \left( \begin{array}{cc}
-1_{9-p} & 0 \\
0 & 1_{p+1}
\end{array}
\right) \ ,\,\, \mand
M_p = \left\{\begin{array}{cc}\frac{\pm i}{(p+1)!}\ga^{i_{1}}\ga^{i_{2}}\ldots \ga^{i_{p+1}}
\eps_{i_{1}\ldots i_{p+1}}\,\,\,\,{\rm for\, p \,even}\\ \frac{\pm 1}{(p+1)!}\ga^{i_{1}}\ga^{i_{2}}\ldots \ga^{i_{p+1}}\ga_{11}
\eps_{i_{1}\ldots i_{p+1}} \,\,\,\,{\rm for\, p \,odd}\end{array}\right.
\non\end{displaymath}
\vskip .2in
Having set the above matrices, we would be able  to make use of  holomorphic part of the two point functions or standard propagators  for  all the fields of $X^{\mu},\psi^\mu, \phi$, as below
\begin{eqnarray}
\lan X^{\mu}(z)X^{\nu}(w)\ran & = & -\frac{\alpha'}{2}\eta^{\mu\nu}\log(z-w) \ , \non \\
\lan \psi^{\mu}(z)\psi^{\nu}(w) \ran & = & -\frac{\alpha'}{2}\eta^{\mu\nu}(z-w)^{-1} \ ,\non \\
\lan\phi(z)\phi(w)\ran & = & -\log(z-w) \ .
\labell{prop2}\end{eqnarray}

By considering the above vertex operators , the S-matrix can be written down as follows 
\beqa
&&\int dx_1 dx_2 dx_4 dx_5  \bar u^{\gamma} u^{\delta}(x_{12}x_{14}x_{15}x_{24}x_{25}x_{45})^{-1/4} 
(P_{-}\fsH_{(n)}M_p)^{\alpha\beta} \nonumber\\&&\times
|x_{12}|^{-2u}|x_{14}x_{15}x_{24}x_{25}|^{u} |x_{45}|^{-2u}<:S_{\al}(x_4): S_{\be}(x_5):S_{\gamma}(x_1):S_{\delta}(x_2):>,
 \nonumber\eeqa
where $x_4=z=x+iy,x_5=\bar z=x-iy$ and $u = -\frac{\alpha'}{2}(k_1+k_2)^2$. The correlation function of four spin operator (with the same chirality) in type IIB is as follows
\beqa
<:S_{\gamma}(x_1):S_{\delta}(x_2):S_{\al}(x_4): S_{\be}(x_5):>&=&\bigg[(\gamma^\mu C)_{\al\be}(\gamma^\mu C)_{\gamma\delta} x_{15} x_{24}-(\gamma^\mu C)_{\gamma\be}(\gamma^\mu C)_{\al\delta} x_{12} x_{45}\bigg]\nonumber\\&&\times
\frac{1}{2(x_{12}x_{14}x_{15}x_{24}x_{25}x_{45})^{-3/4}}
\label{ee12}\eeqa

If we apply the above correlator to the amplitude, then we could readily check the $SL(2,R)$ invariance of the S-matrix. To actually remove the volume of conformal killing group we choose the gauge fixing as  $(x_1,x_2,z,\bar z)=(x,-x,i,-i)$ and the Jacobian becomes $J=-2i(1+x^2)$. Having set the above gauge fixing, we find out the final form of S-matrix as follows 
 
\beqa
&&\int_{-\infty}^{\infty} dx (2x)^{-2u} \bigg( (\gamma^\mu C)_{\al\be}(\gamma^\mu C)_{\gamma\delta} (\frac{-1+x^2}{2x}+i)+2i(\gamma^\mu C)_{\gamma\be}(\gamma^\mu C)_{\al\delta}\bigg)
(1+x^{2})^{-1+2u} \nonumber\\&&\times\frac{(2i)^{-2u}}{2}(P_{-}\fsH_{(n)}M_p)^{\alpha\beta} \bar u^{\gamma} u^{\delta}
,\nonumber\eeqa
 Momentum conservation on brane's world volume is $k_1^{a} + k_2^{a} + p^{a} =0$, obviously the first term in the amplitude has zero contribution to the S-matrix as the integrand is odd function and the interval is symmetric. The second and third term of the above amplitude are contact interactions and the integral for the 2nd term (likewise the result for 3rd term) can be derived. The ultimate result for the amplitude becomes
\beqa
{\cal A}^{\bar\Psi\Psi,RR}_{IIB} &=&(\mu_p/2)2^{-2u-1}\sqrt{\pi} \frac{\Ga(-u+1/2)}{\Ga(1-u)} \bar u^{\gamma} (\gamma^\mu C)_{\gamma\delta}u^{\delta}
 \Tr
(P_{-}\fsH_{(n)}M_p\gamma^{\mu})\Tr(\lam_1\lam_2) \labell{amp383}\ .
\eeqa

where   $ (\mu_p/2)$ is a normalisation constant and $  \mu_p $ is  RR's brane  charge. If  $\mu$ picks the world volume indices up ($\mu=a$), we then  get to know that the trace is non-zero for  $p= n$ case and it can be extracted out  as 
\beqa
\Tr\bigg(\fsH_{(n)}M_p
\gamma^a\bigg)\delta_{p,n}&=&\pm\frac{32}{(p)!}\eps^{a_{0}\cdots a_{p-1}a}H_{a_{0}\cdots a_{p-1}}
  \delta_{p,n}\nonumber\eeqa
  Notice that inside the trace the term including $\gamma^{11}$ confirms that all results are being held for the following as well
\beqa
  p>3 , H_n=*H_{10-n} , n\geq 5.
  \nonumber\eeqa
 The expansion is low energy expansion which is $u=-p^ap_a \rightarrow 0$,  expanding the Gamma functions inside the the amplitude we then would clearly gain all infinite  higher derivative corrections of a field strength RR potential $p$ form field and two fermions with the same chirality. The momentum expansion is 
\beqa
2^{(-2u-1)}\sqrt{\pi}\frac{\Ga(-u+1/2)}{\Ga(1-u)}
 &=&\pi \sum_{m=-1}^{\infty}c_m(u)^{m+1}
\ .\labell{taylor61}\nonumber
\eeqa
 with some of the coefficients to be 
\beqa
c_{-1}=1/2,c_0=0,c_1=\frac{\pi^2}{12},c_2=\z(3),\nonumber\\
c_3=\frac{19\pi^4}{720}, c_4=\frac{\pi^2}{6}\z(3)+3z(5),\eeqa

 Note that these coefficients are different from the coefficients that have shown up in the expansion of non-BPS amplitude of an RR, a tachyon and an scalar field $CT\phi$. 
 This clearly confirms that the normalisation of WZ action of BPS branes is different from non-BPS branes. The first term in the expansion is contact interaction and can be produced by the following supersymmetric Wess-Zumino coupling
 \beqa
(2\pi\alpha')\frac{\mu_p}{p!}\epsilon^{a_{0}...a_{p-1}a}H_{a_{0}...a_{p-1}}\bar\Psi^{\gamma} (\gamma^a)_{\gamma\delta} \Psi^{\delta}
\nonumber\eeqa
 consequently all order $\alpha'$ higher derivative corrections can be constructed by applying the proper higher derivative corrections to the above EFT coupling and also by comparing each term with its string theory elements so that the closed form of corrections to all orders in IIB is demonstrated by 
 \beqa
(2\pi\alpha')\frac{\mu_p}{p!}\epsilon^{a_{0}...a_{p-1}a}H_{a_{0}...a_{p-1}}\sum_{m=-1}^{\infty}c_m(\alpha')^{m+1}D_{a_1}\cdots D_{a_{m+1}}\bar\Psi^{\gamma} (\gamma^a)_{\gamma\delta} D^{a_1}...D^{a_{m+1}}\Psi^{\delta}
\nonumber\eeqa
 Note that if $\mu$ takes the value from transverse directions ($\mu=i$), then the amplitude is non zero for $n=p+2$ case. One can show that all the higher derivative corrections can be looked for by the following coupling in an EFT 
  \beqa
(2\pi\alpha')\frac{\mu_p}{(p+1)!}\epsilon^{a_{0}...a_{p}}H^{i}_{a_{0}...a_{p}}\sum_{m=-1}^{\infty}c_m(\alpha')^{m+1}D_{a_1}\cdots D_{a_{m+1}}\bar\Psi^{\gamma} (\gamma^i)_{\gamma\delta} D^{a_1}...D^{a_{m+1}}\Psi^{\delta}
\label{esi987}\eeqa

Note that the computations in this section give only the derivative pieces of \reef{esi987} and not the full covariant derivatives and indeed using gauge invariance one  can covariantize the action.  It is worthwhile to point out a remark. Likewise the result for supersymmetric amplitude , the result and corrections for asymmetric amplitude of $<V_{\bar\Psi}^{(1/2)}(x_1)V_{\Psi}^{(-1/2)}(x_2) V_{RR}^{(-2)}(z,\bar{z})>$ are also the same, as there is no picture dependence of supersymmetric fermionic amplitudes. Let us look at its IIA version to explore whether or not the structures and coefficients of $\alpha'$ corrections of IIA are different from their IIB ones.
\subsection{ The entire form of $\bigg( RR \bar\Psi^{\dga} \Psi^{\dde}\bigg)$ in type IIA  }

In this section we would like to see whether or not there are some singularities for a particular one RR and two fermion fields with different chirality in IIA. More crucially,  our aim is to construct all order $\alpha'$ higher derivative corrections to these elements as well. There is no issue of picture dependence for mixed closed RR and fermion fields,  hence, for simplicity we just deal with RR in its symmetric picture.

The correlation function of four spin operators with different chirality in IIA was given  by \cite{Friedan:1985ge,Hatefi:2014saa} as follows
\beqa
  <S_{\alpha}(x_4)S_{\beta}(x_5)S^{\dga}(x_1)S^{\dde}(x_2)>= \bigg(\frac{x_{45}x_{12}}{x_{41}x_{42}x_{51}x_{52}}\bigg)^{1/4}
  \bigg[\frac{C_{\alpha}^{\dde}C_{\beta}^{\dga}} {x_{42}x_{51}}-\frac{C_{\alpha}^{\dga}C_{\beta}^{\dde}} {x_{41}x_{52}}+\frac{1}{2}\frac{(\gamma^\mu C)_{\alpha\beta}\,(\bar\gamma_\mu C)^{\dga\dde}} {x_{45}x_{12}}\bigg]\label{ham}\eeqa
Having replaced the above correlation function  into the amplitude, we would be able to obtain the final form of  IIA amplitude  in a manifest way as follows
\beqa {\cal A}& = &
 \int
 dx_{1}dx_{2}dx_{4} dx_{5}\,
(P_{-}\fsH_{(n)}M_p)^{\alpha\beta} \bu ^{\dga} u^{\dde}  (x_{12}x_{14}x_{15}x_{24}x_{25}x_{45})^{-1/4}|x_{12}|^{-2u}|x_{14}x_{15}x_{24}x_{25}|^{u}|x_{45}|^{-2u}
\nonumber\\&&\times \bigg(\frac{x_{45}x_{12}}{x_{41}x_{42}x_{51}x_{52}}\bigg)^{1/4}
  \bigg[\frac{C_{\alpha}^{\dde}C_{\beta}^{\dga}} {x_{42}x_{51}}-\frac{C_{\alpha}^{\dga}C_{\beta}^{\dde}} {x_{41}x_{52}}+\frac{1}{2}\frac{(\gamma^\mu C)_{\alpha\beta}\,(\bar\gamma_\mu C)^{\dga\dde}} {x_{45}x_{12}}\bigg]  \labell{125}\eeqa
  
 We used the same gauge fixing as in IIB one and the final form of amplitude can be packed as follows
  \beqa {\cal A}& = &  (P_{-}\fsH_{(n)}M_p)^{\alpha\beta} \bu^{\dga} u^{\dde}   \int_{-\infty}^{\infty}
 dx (2x)^{-2u} (x^2+1)^{2u} 2^{-2u+1}
\nonumber\\&&\times   \bigg[ (C_{\alpha}^{\dde}C_{\beta}^{\dga}-C_{\alpha}^{\dga}C_{\beta}^{\dde}) (x^2-1)+2ix(C_{\alpha}^{\dde}C_{\beta}^{\dga}+C_{\alpha}^{\dga}C_{\beta}^{\dde})
-\frac{1} {8ix} (\gamma^\mu C)_{\alpha\beta}\,(\bar\gamma_\mu C)^{\dga\dde} \bigg] 
\labell{amp3q},\eeqa
where the  2nd and 3rd term have zero contribution to our S-matrix. Having evaluated the integrals, the final form of the amplitude in IIA would be given by
  \beqa
{\cal A}^{\bar\Psi\Psi,RR}_{IIA} &=&  \frac{\mu_p}{32} (P_{-}\fsH_{(n)}M_p)^{\alpha\beta}2^{-2u+3}\sqrt{\pi} \frac{\Ga(-u-3/2)}{\Ga(-u)} \bu ^{\dga} u^{\dde} (C_{\alpha}^{\dde}C_{\beta}^{\dga}-C_{\alpha}^{\dga}C_{\beta}^{\dde})\labell{amp383}\ .
\eeqa
The trace is non-zero for  $p+1= n$ case, and it can be found out through the way we did in the previous section. The expansion is \beqa
2^{(-2u)}\sqrt{\pi}\frac{\Ga(-u-3/2)}{\Ga(-u)}
 &=&\pi \sum_{m=-1}^{\infty}c_m(u)^{m+1}
\ .\labell{taylor61}\nonumber
\eeqa
 with the following coefficients
\beqa
c_{-1}=0,c_0=-\frac{4}{3},c_1=\frac{32}{9},c_2=\frac{-2}{27}(3\pi^2+104),\nonumber\\
c_3=\frac{-8}{81}(-6\pi^2+27\z(3)-160),\eeqa
It is now clarified that  these coefficients are different from the coefficients that have appeared  in the expansion of  $C\bar\Psi\Psi$ of type IIB of the previous section. The first contact interaction can be generated by the following supersymmetric Wess-Zumino coupling of IIA
 \beqa
\frac{2\pi\alpha'^2c_0\mu_p}{(p+1)!} D^a \bar\Psi^{\dga} D_a\Psi^{\dde}\epsilon^{a_{0}...a_{p}}H_{a_{0}...a_{p}}
\nonumber\eeqa
and all order $\alpha'$ higher derivative corrections can be derived  by comparing with string amplitudes as follows
  \beqa
\frac{2\pi\alpha'^2\mu_p}{(p+1)!} \sum_{m=-1}^{\infty}c_m(\alpha')^{m+1}D_a D_{a_1}\cdots D_{a_{m+1}}\bar\Psi^{\dga}  D^aD^{a_1}...D^{a_{m+1}}\Psi^{\dde}  \epsilon^{a_{0}...a_{p}}H_{a_{0}...a_{p}}\label{ppo01}
\eeqa

where the first correction for IIA couplings (unlike IIB) appears to be at $\alpha'^2$ order. As it is evident not only the structures but also the coefficients of the corrections of IIA\reef{ppo01} are very different from IIB ones \reef{esi987}. The reasons and intuitions for this conclusion are as follows.  Indeed not only $\alpha'$ corrections keep changing at each order but also there is no definite rule for finding $\alpha'$ corrections of fermionic couplings, note also that they obviously couple to different RR forms as well.

\section{ RR couplings  of type IIB and IIA}
In this section we  would like to use CFT to build not only singularities but also all the infinite contact interactions  as well as $\alpha'$ corrections of two closed string RR at disk level. Clearly all spin operators in type IIB carry the same chirality, hence,  this four point function in IIB of BPS branes $ \lan
V_{C}^{(-1)}(x_1,x_2) V_{C}^{(-1)}(x_4,x_5)\ran$ can be found by exploring all the  correlation functions  
where  C-vertex has already been given. The definitions for projection operator and the other matrices as well as notations kept held here as well.
\vskip.1in

 Consider $n_o$ open strings and $n_c$ closed strings, we then have $(n_o+n_c)(n_o+n_c-3)/2$ independent variables from the tangent to the brane momenta.
This takes into account the momentum conservation constrain along the
brane, also we will have  $n_c(n_c-1)/2$ variables of the $p_iNp_j$  so that $i\neq j$
and $n_c$ variables of the type $p_iNp_i=-p_iVp_i$ since $p^2=0$.

Therefore, in general we have 
$(n_o+n_c)(n_o+n_c-3)/2+ n_c(n_c+1)/2$ independent variables.  Indeed we might think of having 3 independent Mandelstam variables for this world sheet four point function of $CC$ amplitude, however, $t+s+u=0$ and therefore u can be removed in terms of s and t.  Thus, we define $s=-\frac{\alpha'}{2}(p_1+D.p_1)^2$ and $t=-\frac{\alpha'}{2}(p_1+p_2)^2$ so that $p_2.D.p_2=p_1.D.p_1$ and $p_1.D.p_2=\frac{s+t}{2}$. The correlation function of four spin operator in IIB has been given in the previous section so the S-matrix is got to be \beqa
&&\int dx_1 dx_2 dx_4 dx_5  (P_{-}\fsH_{(1n)}M_p)^{\alpha\beta}(P_{-}\fsH_{(2n)}M_p)^{\gamma\delta}(x_{12}x_{14}x_{15}x_{24}x_{25}x_{45})^{-1}
 \nonumber\\&&\times \frac{1}{2} \bigg[(\gamma^\mu C)_{\al\be}(\gamma^\mu C)_{\gamma\delta} x_{15} x_{24}-(\gamma^\mu C)_{\gamma\be}(\gamma^\mu C)_{\al\delta} x_{12} x_{45}\bigg]\nonumber\\&&\times
|x_{12}x_{45}|^{-s/2}|x_{14}x_{25}|^{-t/2}|x_{15}x_{24}|^{(s+t)/2}\nonumber\eeqa
 
We could  check the $SL(2,R)$ invariance of the S-matrix as well. To actually remove the volume of conformal killing group, we have chosen the gauge fixing as  follows
\beqa
(x_1,x_2,x_4,x_5)=(iy,-iy,i,-i), Jacobian= -2i(1-y^2),   0\leq y\leq 1 \nonumber\eeqa
Indeed we map it to disk, setting the above gauge fixing, we reveal the final form of S-matrix as follows  
\beqa
{\cal A}^{C^{-1}C^{-1}}_{IIB} &=&\int_{0}^{1} dy (y)^{-s/2-1}(1-y)^{-t-1}(1+y)^{s+t-1} \bigg( (y+1)^2 (\gamma^\mu C)_{\al\be}(\gamma^\mu C)_{\gamma\delta} \nonumber\\&&+4y(\gamma^\mu C)_{\gamma\be}(\gamma^\mu C)_{\al\delta}\bigg)
2^{-s-2}i(P_{-}\fsH_{(1n)}M_p)^{\alpha\beta}(P_{-}\fsH_{(2n)}M_p)^{\gamma\delta} 
,\nonumber\eeqa
Now in order to actually obtain the solutions for integrals in terms of Euler functions, the best way is to deal with the following transformation 
\beqa
y=\frac{1-x^{1/2}}{1+x^{1/2}}
\nonumber
\eeqa
which maps all the integrals to radial integrals on the unit disk. So the whole above S-matrix will be divided to two distinct parts and the solutions after coordinate transformation will be given by
\beqa
I_1=-\int_{0}^{1} dx (x)^{-t/2-1}(1-x)^{-s/2-1}=- \frac{\Ga(-s/2)\Ga(-t/2)}{\Ga(-s/2-t/2)} \nonumber\eeqa
\beqa
I_2=-\int_{0}^{1} dx (x)^{-t/2-1}(1-x)^{-s/2}=- \frac{\Ga(-s/2+1)\Ga(-t/2)}{\Ga(-s/2-t/2+1)}\nonumber\eeqa

Since the expansion is low energy expansion, one can send off $\alpha' $ to zero and start discovering, singularity and  contact terms related to  two closed string RR of type IIB. The expansions of $I_1$ and $I_2$ are accordingly
 \beqa
 I_1&=&\frac{2(s+t)}{ts}-\frac{\pi^2(s+t)}{12}-\frac{1}{4}\z(3)(s+t)^2-\frac{1}{2880}(s+t)\pi^4 (4s^2+st+4t^2)+...\nonumber\\
 I_2&=&\frac{2}{t}-\frac{\pi^2s}{12}-\frac{1}{4}\z(3)s(s+t)-\frac{1}{2880}\pi^4 s(4s^2+st+4t^2)+...\nonumber\eeqa
Note that one could write down the compact form of the above series, for instance the closed form of $I_2$ is given by
\beqa
I_2=\frac{2}{t}-s\sum_{n,m=0}^{\infty} h_{n,m} (ts)^n(t+s)^m
\nonumber\eeqa
Having extracted the traces and further simplifications the ultimate and closed form of two closed string RR amplitude in IIB  to all orders in  $\alpha'$ would be written down by 
\beqa
{\cal A}^{CC}_{IIB} &=& \frac{i\mu_{1p}\mu_{2p}}{ p!p!}
\bigg(\frac{2}{s}-t\sum_{n,m=0}^{\infty} h_{n,m} (ts)^n(t+s)^m\bigg)\nonumber\\&&\times
\eps_{1}^{a_{0}\cdots a_{p-1}a}H_{1a_{0}\cdots a_{p-1}}\eps_{2}^{a_{0}\cdots a_{p-1}a}H_{2a_{0}\cdots a_{p-1}}
\label{esi99}\eeqa
where   $ (\mu_{1p},\mu_{2p})$ are the first and the second RR charge of branes. We have chosen $\mu$ to take values on  world volume directions ($\mu=a$), so that all the traces are non-zero for  $p= n$. The presence of the first singularity clearly shows that we do have just a simple  gauge field singularity that propagates between two $p$-form closed string RR as well as all infinite $\alpha'$ higher derivative corrections to two RR's of IIB.

\vskip.1in

Note that if  $\mu$ takes value on transverse directions ($\mu=i$), then traces  make sense for  $p+2= n$ case and evidently we would get just first simple scalar field singularity structure that propagates between two $p+1$-form closed string RR as well as all the same (but with different H) infinite $\alpha'$ higher derivative corrections to two RR's as follows
\beqa
{\cal A}^{CC}_{IIB} &=&\frac{i\mu_{1p}\mu_{2p}}{(p+1)!(p+1)!}
\eps_{1}^{a_{0}\cdots a_{p}}H^i_{1a_{0}\cdots a_{p}}\eps_{2}^{a_{0}\cdots a_{p}}H^i_{2a_{0}\cdots a_{p}}\bigg(\frac{2}{s}-t\sum_{n,m=0}^{\infty} h_{n,m} (ts)^n(t+s)^m\bigg)
\nonumber\eeqa
We normalised the amplitude by $\frac{1}{2^6 }$,
let us now reconstruct the simple gauge (scalar) pole and continue explaining all order $\alpha'$ corrections.
\beqa
{\cal A}&=&V^a_{\alpha}(C_{1p-1},A)G^{ab}_{\alpha\beta}(A)V^b_{\beta}(C_{2p-1},A),\labell{amp44390}
\eeqa
and scalar pole by 
\beqa
{\cal A}&=&V^i_{\alpha}(C_{1p+1},\phi)G^{ij}_{\alpha\beta}(\phi)V^j_{\beta}(C_{2p+1},\phi),\labell{amp4439055}
\eeqa
where$V^a_{\alpha}(C_{1p-1},A)$  is obtained from the Chern-Simons coupling as 
\beqa
i(2\pi\alpha')\mu_{1p}\int_{\Sigma_{p+1}}C_{1p-1}\wedge F\label{esi982}\eeqa

 and accordingly $V^i_{\alpha}(C_{1p+1},\phi)$  can be gained  from the Taylor expansion of scalar field with mixed RR in an  effective field theory coupling  
\beqa
i(2\pi\alpha')\mu_{1p}\int_{\Sigma_{p+1}} \partial^i C_{1p+1}\phi_i \label{esi984}\eeqa
 
The gauge field and scalar field propagators are produced from their kinetic term in DBI action as $(2\pi\alpha')^2 F_{ab} F^{ab} $ and $\frac{(2\pi\alpha')^2}{2}  \Tr(D_a\phi^i D^a\phi_i)$. We do not have any other gauge (scalar) poles , simply because the kinetic terms of gauge fields and scalars have already been fixed and there are no correction to them any more. Notice that there is no correction to the Chern-Simons and (WZ) couplings  of an RR and a gauge (scalar) field either. The vertices could be easily established by
\beqa
V^a_{\alpha}(C_{1p-1},A)&=&i(2\pi\alpha')\frac{\mu_{1p}}{p!} \eps_{1}^{a_{0}\cdots a_{p-1}a}H_{1a_{0}\cdots a_{p-1}}
  \Tr( \lambda_\alpha)
\nonumber\\
V^i_{\alpha}(C_{1p+1},\phi)&=&i(2\pi\alpha')\frac{\mu_{1p}}{(p+1)!} \eps_{1}^{a_{0}\cdots a_{p}}H^{i}_{1a_{0}\cdots a_{p}}
  \Tr( \lambda_\alpha)
\nonumber\\
G_{\alpha\beta}^{ab}(A)&=&\frac{-1}{(2\pi\alpha')^2}\frac{\delta^{ab}
\delta_{\alpha\beta}}{k^2}\,\,\, ,\nonumber\\
V^b_{\beta}(C_{2p-1},A)&=&i(2\pi\alpha')\frac{\mu_{2p}}{p!} \eps_{2}^{a_{0}\cdots a_{p-1}b}H_{2a_{0}\cdots a_{p-1}}
  \Tr( \lambda_\beta)
\label{ver138}
\eeqa
 $k^2=-(p_1+D.p_1)^2=-s$  should also be substituted in the propagator.  
  
Replacing \reef{ver138}  into \reef{amp44390} and \reef{amp4439055} appropriately, we are precisely able to find out the gauge field (scalar field) singularity of string amplitude in an EFT.  Given the closed and all order form of the amplitude in \reef{esi99}, now one starts to apply properly all order $\alpha'$ higher derivative corrections to two closed string RR of type IIB as follows
\beqa
&&\sum_{n,m=0}^{\infty} h_{n,m} (\alpha')^{m+2n+1}  (D_aD_a)^m\bigg(D_{a_{1}}...D_{a_{n+1}} (D_bD_b)^n C_{1a_0\cdots a_{p-2}}  D^{a_{1}}...D^{a_{n+1}} C_{2a_0\cdots a_{p-2}}\bigg)\nonumber\\&&\times
\frac{\mu_{1p} \mu_{2p}}{(p-1)!(p-1)!}\eps_{1}^{a_{0}\cdots a_{p-2}a}\eps_{2}^{a_{0}\cdots a_{p-2}a}\label{5gh}
\eeqa
Let us find RR couplings and their corrections in type IIA. Clearly here just the correlation function of four spin operator gets changed and the same definitions for Mandelstam variables as well as the same notations are being held. Therefore  the amplitude in IIA is given by
 \beqa
&&\int dx_1 dx_2 dx_4 dx_5  (P_{-}\fsH_{(1n)}M_p)^{\alpha\beta}(P_{-}\fsH_{(2n)}M_p)^{\dga\dde}  (x_{12}x_{14}x_{15}x_{24}x_{25}x_{45})^{-1/4} \nonumber\\&&\times \bigg(\frac{x_{45}x_{12}}{x_{41}x_{42}x_{51}x_{52}}\bigg)^{1/4}
  \bigg[\frac{C_{\alpha}^{\dde}C_{\beta}^{\dga}} {x_{42}x_{51}}-\frac{C_{\alpha}^{\dga}C_{\beta}^{\dde}} {x_{41}x_{52}}+\frac{1}{2}\frac{(\gamma^\mu C)_{\alpha\beta}\,(\bar\gamma_\mu C)^{\dga\dde}} {x_{45}x_{12}}\bigg] 
\nonumber\\&&\times
|x_{12}x_{45}|^{-s/2}|x_{14}x_{25}|^{-t/2}|x_{15}x_{24}|^{(s+t)/2} \labell{ee125}\eeqa
 We carry out the same  gauge fixing as  $(x_1,x_2,x_4,x_5)=(iy,-iy,i,-i), J= -2i(1-y^2)$ and eventually the amplitude gets reduced to
\beqa
{\cal A}^{C^{-1}C^{-1}}_{IIA} &=&-2i\int_{0}^{1} dy (y)^{-s/2}(1-y)^{-t}(1+y)^{s+t} \bigg[ \frac{C_{\alpha}^{\dde}C_{\beta}^{\dga}}{(y+1)^2}+\frac{C_{\alpha}^{\dga}C_{\beta}^{\dde}}{(1-y)^{2}}-\frac{1} {8y} (\gamma^\mu C)_{\alpha\beta}\,(\bar\gamma_\mu C)^{\dga\dde} \bigg] \nonumber\\&&\times
2^{-s}(P_{-}\fsH_{(1n)}M_p)^{\alpha\beta}(P_{-}\fsH_{(2n)}M_p)^{\dga\dde} 
\labell{amp3qgg},\eeqa
We also map  the integrals to radial integrals on the unit disk and use the same change of variable as $y=\frac{1-x^{1/2}}{1+x^{1/2}}$. Hence, the whole  S-matrix is going to be  divided to three different parts. One needs to evaluate the integrals where we just illustrate the ultimate result as follows  
\beqa 
I_1&=&-2^{-s}\int_{0}^{1} 2^{s-2}dx (x)^{-t/2-1/2}(1-x)^{-s/2}=- \frac{\Ga(-s/2+1)\Ga(-t/2+1/2)}{4\Ga(-s/2-t/2+3/2)} \nonumber\\
 I_2&=&-2^{s-2}2^{-s}\int_{0}^{1} dx (x)^{-t/2-3/2}(1-x)^{-s/2}=- \frac{\Ga(-s/2+1)\Ga(-t/2-1/2)}{4\Ga(-s/2-t/2+1/2)}\nonumber\\
 I_3&=&-2^{s-3}2^{-s}\int_{0}^{1} dx (x)^{-t/2-1/2}(1-x)^{-s/2-1}=- \frac{\Ga(-s/2)\Ga(-t/2+1/2)}{8\Ga(-s/2-t/2+1/2)}\nonumber\eeqa
The expansion is again low energy expansion so  we send  $\alpha' $ to zero and start to reveal singularity and  contact terms related to  two closed string RR of type IIA. Obviously, $I_1,I_2$  just include all the contact interactions, meanwhile $I_3$ has just a simple s-channel pole and some of the expansions are given by
 \beqa
 I_2&=&\frac{-1}{4}\bigg(-2+2ln2 s+2t+s^2(\frac{\pi^2}{12}-2ln2)+st(-2ln2+\frac{\pi^2 }{4})
+...\bigg),\nonumber\\
 I_3&=&\frac{-1}{8}\bigg(\frac{-2}{s}+2ln2 +\frac{\pi^2s}{12}-2ln2 s+\frac{\pi^2t}{4}+s^2(\frac{1}{2}\z(3)+ln2 -\frac{\pi^2 ln2}{12})\nonumber\\&&+st(\frac{7}{4}\z(3)-\frac{\pi^2 ln2 }{4})+\frac{7}{4}\z(3)t^2+....\bigg)\nonumber\eeqa
The closed form of $I_3$ is given by
\beqa
I_3=\frac{-1}{8}\bigg(\frac{-2}{s}+\sum_{n,m=0}^{\infty} l_{n,m} s^n t^m\bigg)
\nonumber\eeqa

Extracting the traces and further simplifications, one might explore  the closed form of the third term of \reef{amp3qgg}  to all orders in $\alpha'$ by the following algebraic function 
\beqa
{\cal A}^{CC}_{IIA} &=& \frac{i\mu_{1p}\mu_{2p}}{ p!p!}
\eps_{1}^{a_{0}\cdots a_{p-1}a}H_{1a_{0}\cdots a_{p-1}}\eps_{2}^{a_{0}\cdots a_{p-1}a}H_{2a_{0}\cdots a_{p-1}}\bigg(-\frac{2}{s}+\sum_{n,m=0}^{\infty} l_{n,m} s^n t^m\bigg)
\label{esi348}\eeqa

We have chosen $\mu$ to take values on  world volume directions ($\mu=a$), thus all the traces are non-zero for  $p= n$. The presence of the first singularity clearly shows that we do have just a simple  gauge field singularity that propagates between two $p$-form closed string RR as well as all infinite $\alpha'$ higher derivative corrections to two RR's of IIA.
Note that if  $\mu$ takes value on transverse directions ($\mu=i$), then the traces will have non vanishing values  for  $p+2= n$ case. Evidently we would also get just a simple scalar field singularity that propagates between two $p+1$-form closed string RR as well as contact terms as follows
\beqa
{\cal A}^{CC}_{IIA} &=&\frac{i\mu_{1p}\mu_{2p}}{(p+1)!(p+1)!}
\eps_{1}^{a_{0}\cdots a_{p}}H^i_{1a_{0}\cdots a_{p}}\eps_{2}^{a_{0}\cdots a_{p}}H^i_{2a_{0}\cdots a_{p}}\bigg(-\frac{2}{s}+\sum_{n,m=0}^{\infty} l_{n,m} s^n t^m\bigg)
\label{bbc1}\eeqa

Looking carefully at \reef{bbc1}, we come to know that unlike the structures of corrections,  the coefficients of all infinite $\alpha'$ higher derivative corrections to two RR's of IIA for this case would be the same as appeared in \reef{esi348}.

It is worth to highlight the fact that both simple s-channel  gauge field and scalar field can be precisely reconstructed in an effective field theory by the same rules of \reef{amp44390} and \reef{amp4439055} appropriately, where all the vertices have also been pointed out in \reef{ver138}. Let us fully address the point of this section, which is finding out not only structures but also compact and the closed form of the coefficients of  all order $\alpha'$ higher derivative corrections of 2 RR of IIA one. Indeed one can start to compare order by order the elements of string amplitude with effective field theory couplings and eventually provide all the corrections involving their structures of IIA as follows
\beqa
&&\sum_{n,m=0}^{\infty} l_{n,m} (\alpha')^{m+n}  \bigg((D^bD_b)^n D_{a_{1}}...D_{a_{m}} C_{1a_0\cdots a_{p-2}}  D^{a_{1}}...D^{a_{m}} C_{2a_0\cdots a_{p-2}}\bigg)
\nonumber\\&&\times
\frac{\mu_{1p} \mu_{2p}}{(p-1)!(p-1)!}\eps_{1}^{a_{0}\cdots a_{p-2}a}\eps_{2}^{a_{0}\cdots a_{p-2}a}\label{5ghllo}
\eeqa
If one considers \reef{bbc1}, then one is able to generate all the corrections including their structures for the only non vanishing particular elements of $n=p+2$  as below
\beqa
&&\sum_{n,m=0}^{\infty} l_{n,m} (\alpha')^{m+n}  \bigg((D^bD_b)^n D_{a_{1}}...D_{a_{m}} C^{i}_{1a_0\cdots a_{p-1}}  D^{a_{1}}...D^{a_{m}} C_{2ia_0\cdots a_{p-1}}\bigg)
\nonumber\\&&\times
\frac{\mu_{1p} \mu_{2p}}{p!p!}\eps_{1}^{a_{0}\cdots a_{p-1}}\eps_{2}^{a_{0}\cdots a_{p-1}}\label{5ghllokio}
\eeqa

\vskip.1in
To  end this section, we provide all the other contact interactions that are produced by $I_1,I_2$ as well. By extracting the related traces and carrying out some further algebraic simplifications, one can establish the closed form of the contact interactions of two closed string RR amplitude to all order $\alpha'$ in IIA   as follows
\beqa
  \frac{i\mu_{1p}\mu_{2p}}{ (p+1)!(p+1)!}
\eps_{1}^{a_{0}\cdots a_{p}}H_{1a_{0}\cdots a_{p}}\eps_{2}^{a_{0}\cdots a_{p}}H_{2a_{0}\cdots a_{p}}
\bigg( \sum_{n,m=0}^{\infty} k_{n,m} s^n t^m\bigg)
\labell{amp3qggmm},\eeqa
 
where some of the coefficients are 
\beqa
k_{0,0}=-4,k_{1,0}=4ln2-2,k_{2,0}=-2ln2-2+\frac{\pi^2}{6},k_{1,1}=\frac{\pi^2}{2}-4,k_{0,2}=-4\nonumber\eeqa
Given the prescription for the corrections, now one explores the leftover $\alpha'$ higher derivative corrections of $I_1,I_2$  to be
\beqa
&&\sum_{n,m=0}^{\infty} k_{n,m} (\alpha')^{m+n}\bigg(D_{a_{1}}...D_{a_{m}} (D_bD_b)^n C_{1a_0\cdots a_{p-1}}  D^{a_{1}}...D^{a_{m}} C_{2a_0\cdots a_{p-1}}\bigg) \nonumber\\&&\times\frac{\mu_{1p} \mu_{2p}}{(p!)^2}
\eps_{1}^{a_{0}\cdots a_{p-1}}\eps_{2}^{a_{0}\cdots a_{p-1}}\label{5ghbbn}
\eeqa
We are now at the steps to conclude. By comparisons of the corrections in both IIB,IIA now  it becomes evident that not only the  coefficients but also the structures of $\alpha'$ corrections of type IIB are quite different from their IIA ones and this fact becomes known by evaluating direct  CFT techniques  and performing all world sheet calculations to all orders.
 \section{Other world-sheet 4 point functions}
 
For warm-up, we start addressing three point function of BPS branes with their restricted Bianchi identities in both symmetric picture of RR and   in terms of potential C-field (antisymmetric picture of RR).  To start with, we highlight the needed vertex operators in both symmetric and asymmetric pictures as follows 
\beqa
 V_\phi^{(-1)}(x)&=&e^{-\phi(x)}\xi_i\psi^i(x)e^{ \alpha'iq\inn X(x)} \nonumber\\
V_A^{(-1)}(x)&=&e^{-\phi(x)}\xi_a\psi^a(x)e^{ \alpha'iq\inn X(x)} \nonumber\\
V_{\phi}^{(-2)}(x) &=&e^{-2\phi(x)}V_{\phi}^{(0)}(x) \nonumber\\
V_{A}^{(0)}(x) &=& \xi_{1a}(\partial^a X(x)+i\alpha'k.\psi\psi^a(x))e^{\alpha'ik.X(x)}\label{d4Vs} \nonumber\\
V_{RR}^{(-2)}(z,\bar{z})&=&(P_{-}\fsC_{(n-1)}M_p)^{\al\be}e^{-3\phi(z)/2} S_{\al}(z)e^{ip\cd X(z)}e^{-\phi(\bar{z})/2} S_{\be}(\bar{z}) e^{ip\cd D \cd X(\bar{z})}
\eeqa

where the C-vertex operator in antisymmetric picture was earlier pointed out in  \cite{Bianchi:1991eu} and  later on was built  in \cite{Liu:2001qa}.  The world-sheet 3-point function of an antisymmetric closed string RR and a gauge field on the whole ten dimensional spacetime can be derived by $ \lan V_A^{(0)}(x)
V_{C}^{(-2)}(z,\bar{z})\ran$, where at disk level the world volume gauge field is located on the boundary , while the closed string would be located in the middle of disk  and on-shell conditions are $k^2=p^2=0, \quad k.\xi_1=0$ .  Carrying out the correlators, one figures out the S-matrix as follows
\beqa
&&\int dx_1  dx_4 dx_5  (P_{-}\fsC_{(n-1)}M_p)^{\alpha\beta}  (x_{45})^{-3/4}  \xi_{1a}  \bigg(-ip^a \frac{x_{45}}{x_{14}x_{15}}+(2i k_{1b})I_2\bigg) |x_{14}x_{15}|^{\frac{\alpha'^2}{2}k_1.p} |x_{45}|^{\frac{\alpha'^2}{4}p.D.p}
 \nonumber\eeqa
where $x_4=z=x+iy,x_5=\bar z=x-iy$. $I_2$ is related to two spinor and a current correlation function that can be accommodated by the Wick-like rule \cite{Liu:2001qa,Hatefi:2010ik} as below
  \beqa
  I_2= <:S_{\al}(x_4): S_{\be}(x_5):\psi^{b}\psi^{a}(x_1):>&=& 2^{-1}(x_{14}x_{15})^{-1}(x_{45})^{-1/4}(\Gamma^{ab} C^{-1})_{\alpha\beta}
  \nonumber\eeqa
We make use of  $(x_1,z,\bar z)=(\infty,i,-i)$ as gauge fixing and the final result for the amplitude in both type IIB,IIA can be written as follows 
\beqa
{\cal A}^{A^{0},C^{-2}} &=& (2i)^{-1}\xi_{1a}\bigg[-ip^a\Tr(P_{-}\fsC_{(n-1)}M_p)+ik_{1b}\Tr(P_{-}\fsC_{(n-1)}M_p\Gamma^{ab})\bigg] \labell{cc}\eeqa

We just hint out to the final result of the same S-matrix in symmetric picture as well
\beqa
{\cal A}^{A^{-1},C^{-1}} &=& 2^{-1/2}(2i)^{-1} \xi_{1a}\Tr(P_{-}\fsH_{(n)}M_p\gamma^a)\labell{cc22}\eeqa

 Extracting the trace, the result for the symmetric S-matrix is given by
\beqa
{\cal A}^{A^{-1},C^{-1}} &=& 2^{-1/2}(2i)^{-1} \xi_{1a}\frac{16}{p!} \epsilon^{a_0...a_{p-1}a}H_{a_0...a_{p-1}}\labell{cc222}\eeqa
 Now if we multiply the amplitude by $2^{-1/2}\pi\mu_p$ 
 then one realizes that the  S-matrix can be regenerated  by \reef{esi982}.
  
If we simultaneously apply  the momentum conservation along the world volume of brane $(k_1+p)^b=0$ and on-shell condition for the gauge field $p^a\xi_{1a}=-k_1.\xi_1=0$ to the first term \reef{cc}, then we come to know that  
 the 1st term of \reef{cc} has no physical contribution to the asymmetric S-matrix. Eventually, if we extract the trace for the 2nd term of 
asymmetric amplitude and apply  $(k_1+p)^b=0 $ relation, we are then  able to precisely reproduce the asymmetric amplitude by 
 the Chern-Simons coupling  as well. Therefore, in order to make sense of non-vanishing asymmetric amplitude, we also come to the conclusion that the following term
\beqa
p_b\epsilon^{a_0...a_{p-2}ab} C_{a_0...a_{p-2}} 
\label{esi12}\eeqa
is non-zero and therefore $p_b\epsilon^{a_0...a_{p-2}ab}$ is non-zero for BPS branes.    
 
  Note that unlike above, for the mixed RR, scalar fields , one needs to explore the restricted Bianchi identity to actually make sense of asymmetric S-Matrix .The symmetric  3 point function of one RR and a scalar field is 
\beqa
{\cal A}^{C^{-1}\phi^{-1}} &=& 2^{-1/2}\Tr
(P_{-}\fsH_{(n)}M_p\gamma^{i})\xi_{1i} \labell{ee}\ .
\eeqa
where the amplitude can be reconstructed in an EFT by 
   $ \mu_p (2\pi\alpha')\int \partial_{i}C_{p+1}\phi^i$ where the Taylor expansion has been used. The S-matrix  in asymmetric picture was found to be
\beqa
{\cal A}^{\phi^{0},C^{-2}} &=& \bigg[-i p^i\Tr(P_{-}\fsC_{(n-1)}M_p)+ik_{1a}\Tr(P_{-}\fsC_{(n-1)}M_p\Gamma^{ia})\bigg]\xi_{1i}\label{rr12}
\eeqa

If we apply  the momentum conservation  to the 2nd term of asymmetric amplitude $k_1^{a} + p^{a} =0$ and then extract its trace we then come to conclusion that, to be able to get to the same result as appeared in symmetric S-matrix in  \reef{ee} the following strong restricted Bianchi identity should be valid for a transverse scalar field  in the presence of RR
   \beqa
   p^a\eps^{a_{0}\cdots a_{p-1}a} =0
   \eeqa
therefore the 2nd term of \reef{rr12} has no contribution to asymmetric S-matrix and the 1st term in \reef{ee} builds exactly the same contact term of 3 point function. Let us deal with 4-point world sheet S-matrix.
 \subsection{ $\lan V_A^{(0)}V_A^{(0)}
V_{RR}^{(-2)}\ran$ to all orders}

The four point function of an antisymmetric RR closed string and two world volume gauge fields or $<V_{A^{0}} V_{A^{0}} V_{C^{-2}} >$ can be obtained by finding the correlators of $\lan V_A^{(0)}(x_1)V_A^{(0)}(x_2) V_{RR}^{(-2)}(z,\bar{z})\ran$.
 Here we have just one Mandelstam variable that can be introduced to be $u=\frac{-\alpha'}{2} (k_1+k_2)^2$ and using Wick theorem,  the  amplitude  is derived  as below
\beqa
\int dx_1 dx_2 dx_4 dx_5  (P_{-}\fsC_{(n-1)}M_p)^{\alpha\beta}  (x_{45})^{-3/4} \xi_{1a}\xi_{2b}  \bigg(I_1+I_2+I_3+I_4\bigg) |x_{12}|^{-2u}|x_{14}x_{15}|^{u} |x_{24}x_{25}|^{u}|x_{45}|^{-2u}
 \nonumber\eeqa
 If we start applying the generalized form of Wick-like rule then we are able to find all the fermionic correlators as
\beqa
 I_1&=& (-\eta^{ab}(x_{12})^{-2}+a_{1a}a_{2b} )  (x_{45})^{-5/4} ( C^{-1})_{\alpha\beta},\nonumber\\&&
a_{1a} = ik_{2a}\bigg[\frac{x_{42}}{x_{12}x_{14}}+\frac{x_{52}}{x_{12}x_{15}}\bigg]\nonumber\\&&
a_{2b}= ik_{1b}\bigg[\frac{x_{14}}{x_{12}x_{24}}+\frac{x_{15}}{x_{12}x_{25}}\bigg]\nonumber\\&&
I_2 =ik_{2d}a_{1a} (x_{24}x_{25})^{-1}  (x_{45})^{-1/4} ( \Gamma^{bd}C^{-1})_{\alpha\beta}\nonumber\\&&
 I_3= ik_{1c}a_{2b}(x_{14}x_{15})^{-1}  (x_{45})^{-1/4} ( \Gamma^{ac}C^{-1})_{\alpha\beta}\nonumber\\&&
 I_4=-k_{1c}k_{2d} (x_{14}x_{15}x_{24}x_{25})^{-1}  (x_{45})^{3/4} \nonumber\\&&\times \bigg[ ( \Gamma^{bdac}C^{-1})_{\alpha\beta}+2\frac{Re[x_{14}x_{25}]}{x_{12}x_{45}}\bigg(\eta^{cd}(\Gamma^{ba}C^{-1})_{\alpha\beta}-\eta^{cb}(\Gamma^{da}C^{-1})_{\alpha\beta}
- \eta^{ad}(\Gamma^{bc}C^{-1})_{\alpha\beta}\nonumber\\&&+\eta^{ab}(\Gamma^{dc}C^{-1})_{\alpha\beta}\bigg)
+4(\frac{Re[x_{14}x_{25}]}{x_{12}x_{45}})^2(-\eta^{ab}\eta^{cd}+\eta^{ad}\eta^{bc})(C^{-1})_{\alpha\beta}\bigg] \label{mion}\eeqa
We wrote all the elements of the amplitude in such a way that, the $SL(2,R)$ invariance of them becomes manifest. Using the  gauge fixing  as  $(x_1,x_2,z,\bar z)=(x,-x,i,-i)$, Jacobian turns out to be  $-2i(1+x^2)$. Lastly, one could gain the final form of S-matrix as

\vskip.2in

 \beqa
 {\cal A}^{A^{0}A^{0}C^{-2}}&=&\xi_{1a}\xi_{2b}\int_{-\infty}^{\infty}dx (1+x^2)^{2u-1} (2x)^{-2u}(2i)^{-2u} (P_{-}\fsC_{(n-1)}M_p)^{\alpha\beta}
 \bigg[\frac{1-x^2}{x}\bigg(k_{2d}k_{2a}(\Gamma^{bd}C^{-1})_{\alpha\beta}\nonumber\\&&
 -k_{1c}k_{1b}(\Gamma^{ac}C^{-1})_{\alpha\beta}
 -k_{1c}k_{2d}(\eta^{cd}(\Gamma^{ba}C^{-1})_{\alpha\beta}-\eta^{cb}(\Gamma^{da}C^{-1})_{\alpha\beta}
- \eta^{ad}(\Gamma^{bc}C^{-1})_{\alpha\beta}\nonumber\\&&+\eta^{ab}(\Gamma^{dc}C^{-1})_{\alpha\beta})\bigg)
+\frac{k_{1b}k_{2a}}{2i} (\frac{1-x^2}{x})^2(C^{-1})_{\alpha\beta}-\eta^{ab}(2x)^{-2}\frac{(x^2+1)^2}{2i}(C^{-1})_{\alpha\beta}
\nonumber\\&&
 -2ik_{1c}k_{2d}\bigg((\Gamma^{bdac}C^{-1})_{\alpha\beta}+4(\frac{1-x^2}{4ix})^2(\eta^{ad}\eta^{bc}-\eta^{ab}\eta^{cd})(C^{-1})_{\alpha\beta}\bigg)\bigg]\label{lop}\eeqa
 \vskip.1in
 
Now if we simplify the amplitude further we then realize that the 7th and 8th terms of the above amplitude will be cancelled by the 10th and 11th terms of \reef{lop} accordingly, meanwhile the 1st up to the 6th term of \reef{lop} have also  zero contribution to asymmetric S-matrix due to the following reason.
Indeed the integrand is odd function while the interval is symmetric and therefore the outcome is zero. The ultimate result for the amplitude in asymmetric picture is given by
\vskip.1in

 \beqa
{\cal A}^{C^{-2}A^0A^{0}}_{1}=\pm \mu_p \pi\frac{8}{(p-2)!}k_{1c} k_{2d}\xi_{1a}\xi_{2b}
\eps^{a_{0}\cdots a_{p-4}bdac}
C_{a_{0}\cdots a_{p-4}}(2)^{-2u}\frac{\pi^{1/2}\Gamma(-u+1/2)}{ \Gamma(-u+1)},\label{111}
\nonumber\eeqa
where  $ \frac{\mu_p  \pi}{4}$ is a normalisation constant. The amplitude is antisymmetric under interchanging the gauge fields, it is non zero for just non abelian case $p=n+2$ and respects the Ward identity . The expansion is low energy expansion , that is , $u=-p_ap^a\rightarrow 0$ and the function is expanded around it to be

\vskip.1in

\beqa
(2)^{-2u}\frac{\pi^{1/2}\Gamma(-u+1/2)}{\Gamma(-u+1)}&=&{\pi} \sum_{n=-1}^{\infty}b_nu^{n+1}.
\nonumber\eeqa
where some of the coefficients $b_n$ are
\beqa 
&&b_{-1}=1,\,b_0=0,\,b_1=\frac{1}{6}\pi^2,\,b_2=2\z(3)\,b_3=\frac{19}{360}\pi^4,\,b_4=\frac{1}{3}(\pi^2\z(3)+18\z(5)).\nonumber\eeqa
The first term in string amplitude can be regenerated by the following Chern-Simons coupling 
\beqa
\frac{1}{2!}\mu_p(2\pi\alpha')^{2}\Tr (C_{p-3}\wedge F\wedge F).\labell{hdervrr}
\eeqa
and all the other contact terms are related to an infinite higher derivative corrections to the above coupling. Thus one can start to apply all order $\alpha'$ corrections to the above coupling and find out the closed form of all order $\alpha'$ higher derivative corrections as follows.
The non-leading terms are corresponded to the higher derivative correction of \reef{hdervrr}. Thus, the corrections to all orders turned out as the closed form to be
\beqa
\frac{1}{2!}\mu_p(2\pi\alpha')^{2}C_{p-3}\wedge \Tr \bigg(\sum_{n=-1}^{\infty} b_n (\alpha')^{n+1} D_{a_0}...D_{a_n} F\wedge D^{a_0}...D^{a_n}F\bigg).\labell{hderv}
\eeqa

 \vskip.2in

 Finally we just illustrate the restricted Bianchi identity for the scalar field that has to be worked out in the presence of an RR and a gauge field. 
 In order to get the consistent  result for the four point function of an RR, a scalar and a gauge field, in \cite{Hatefi:2015gwa} we have derived all possible ways of distributing super ghost charge and explored the results as follows 
 
 \vskip.2in
 
  \beqa
 {\cal A}^{\phi^{0}A^{0}C^{-2}}&=&-\xi_{1i}\xi_{2a} 2ik_{2c}p^i(2i)^{-2u} \Tr(P_{-}\fsC_{(n-1)}M_p \Gamma^{ac})\pi^{1/2}\frac{\Ga(-u+1/2)}{
 \Ga(1-u)}
  \label{lopbb4568}\eeqa
while with symmetric case the result gets deduced to \beqa
 {\cal A}^{\phi^{-1}A^{0}C^{-1}}&=&2^{-3/2}\xi_{1i}\xi_{2a}\int_{-\infty}^{\infty}dx (1+x^2)^{2u-1} (2x)^{-2u}
 \bigg[-2ik_{2b}\Tr(P_{-}\fsH_{(n)}M_p\Gamma^{abi})\bigg]\nonumber\eeqa
 \vskip.1in
 
and also the other case ${\cal A}^{\phi^{0}A^{-1}C^{-1}}$ would have become
 \beqa
 2^{1/2}\xi_{1i}\xi_{2a}\int_{-\infty}^{\infty}dx (1+x^2)^{2u-1} (2x)^{-2u}
 \bigg[k_{1b}\Tr(P_{-}\fsH_{(n)}M_p\Gamma^{bai})-p^i\Tr(P_{-}\fsH_{(n)}M_p\gamma^{a})\bigg]\label{mty}\eeqa
 
 \vskip.2in

Suppose we apply momentum conservation to the 1st term of  \reef{mty}, and try to make use of the following restricted Bianchi identity
\vskip.2in

 \beqa
 p_b\eps^{a_{0}\cdots a_{p-2}ba} H^i_{a_{0}\cdots a_{p-2}}+p^i\eps^{a_{0}\cdots a_{p-1}a} H_{a_{0}\cdots a_{p-1}} &=&0
 \label{majid}\eeqa
 
 \vskip.2in
 
By doing so, we are precisely able to remove the 2nd term \reef{mty}, more significantly, the derivation of effective action as  \reef{lopbb4568} is held. On the other hand, from the effective field theory side, one can start to construct all order $\alpha'$ higher derivative corrections of an RR, a gauge field and an scalar field in both IIB,IIA  through applying the same pattern (as discussed for an RR and 2-gauge fields). Hence,  all order higher derivative corrections can be constructed out without any ambiguity  by the following coupling 
\vskip.1in
 \beqa  
 (2\pi\alpha')^2\mu_p \sum_{n=-1}^{\infty} b_n (\alpha')^{n+1}\int _{H^{+}}\partial_{i}C_{p-1}\wedge D_{a_0}...D_{a_n}F D^{a_0}...D^{a_n} \phi^i \label{mnb}\eeqa
 where the mixed combination of Taylor expansion and Chern-Simons coupling was made.
\section{Conclusion}
We calculated all three and four point couplings of BPS $D_{p}$-branes for all different cases, including  an RR and two fermion fields with the same or different chirality of IIB and IIA , as well as  two closed string RR and an RR and two gauge (scalar) fields in asymmetric case. Their all order $\alpha'$ higher derivative corrections have also been explored. We also obtained the closed form of  supersymmetric Wess-Zumino (WZ) actions, clarifying that not only the structures of $\alpha'$ corrections but also their coefficients of IIB are quite different from their IIA ones.  Eventually,  we made some remarks on the restricted Bianchi identities for several supersymmetric amplitudes of different field content.
\section*{Acknowledgements}
The author would like to thank P. Anastasopoulos, N.Arkani-Hamed, C. Bachas, Massimo. Bianchi, M.Douglas, C.Hull, H. Steinacker, H. Skarke, R.Unge, L. Mason, K. Narain, C. Nunez, C. Papageorgakis, T.Wrase and D.Young for discussions. He would also like to specially thank L. Alvarez-Gaume for many discussions and supports. This paper was initiated during my 2nd post doc at Queen Mary and I deeply thank QMUL, Oxford, and Swansea Universities for the hospitality. This work is supported by FWF project P26731-N27.  

  \end{document}